\documentstyle[aps,prl,epsfig,floats]{revtex}

\begin{document}

\draft
\wideabs{

\title{Dual proximity effect near superconductor-insulator transition}

\author{Hyok-Jon Kwon }
\address{Department of Physics and Center for Superconductivity
Research, University of Maryland, College Park, MD 20742-4111}
\date{May 10, 2001}

\maketitle
\begin{abstract}
We show that quantum vortex-loop proliferation may be one possible
explanation of the 
super-long-range proximity effect observed in an insulating
underdoped cuprate, using the dual theory of quantum vortices. As a
test of this scenario, we propose that 
a dual proximity effect experiment can confirm the
superfluid motion of the quantum vortices in the vortex-proliferated
insulator and can measure the divergent correlation lengths near the
superconductor-insulator transition.
 \end{abstract}

}

Cuprate superconducting materials exhibit a rich structure of phases
depending on the charge-carrier doping and the temperature.  Most
notably, they exhibit superconducting, antiferromagnetic (AF) Mott-insulating,
and $T>T_c$ strange metallic phases.  To this date, however, the nature
of the intervening phases between the superconducting state and the AF
Mott-insulating state is not understood\cite{pseudogap}.  In attempts to
understand this region of the phase diagram, there have been a number
of theoretical proposals on the mechanism of destruction of
superconductivity such as the superconductor-insulator
transition into the quantum-disordered insulator with vortex-loop
proliferation\cite{SIT1,SIT2,Kim,Kwon}, superconductor into AF insulator
transition with the SO(5)-symmetric theory\cite{SO5}, transition into
novel ground states which are possible in two-dimensional (2D) doped Mott
insulators\cite{spinon,Kim}, or striped
phases\cite{stripes}. Therefore, it would be
desirable to devise experiments which can shed light on the
nature of the underdoped insulating state.

In an experiment performed by Decca {\it et al}.\cite{Decca}, Josephson
junctions of superconductor-insulator-superconductor (SIS) type were
made by generating photo-induced superconducting electrodes  on
insulating underdoped thin films of $\rm YBa_2Cu_3O_{
6+\it x}$ family.  Surprisingly, they observed the Josephson effect 
with an invariant product of the Josephson critical current and the
normal states junction resistance for
separations as large as 100 nm between superconducting electrodes, much
larger than the superconducting coherence length $\xi_0\sim 1$ nm. This
finding indicates enhancement of the effective correlation length
($\xi_c$) of the superconducting order parameter in the insulator, which
can be explained by the proximity to a continuous quantum phase
transition. For instance, the SO(5) theory is one possible scenario which
predicts such an effect\cite{SO5Jo}. In this
paper, as an alternative scenario, we will demonstrate that the vortex
proliferation does provide a long correlation length scale near  SIT,
which may explain the super-long-range proximity effect in the underdoped
insulator. In fact,  experiments on the optical
conductivity\cite{Corson} and the thermal Nernst effects\cite{Xu}
provided evidences  for strong thermal fluctuations of unbound
vortices over a wide range of temperatures above $T_c$ in the underdoped
pseudogap phase, which suggest a Kosterlitz-Thouless like transition with
a broad phase fluctuation regime\cite{Emery}. In addition, an experiment on
current-voltage characteristic of superconducting
Bi2212 shows indications of a large
density of quantum vortex pairs well below $T_c$ and far away from
SIT\cite{Chiaverini}. Therefore, given the indications of strong
quantum/thermal phase fluctuations in cuprates, vortex-proliferation is a
prospective candidate scenario of SIT.

We also propose an
experiment which can detect a qualitatively distinct effect of Josephson
tunneling of superfluid vortices in an insulator-superconductor-insulator
(ISI) junction. If the SIT is induced by the vortex proliferation, 
we expect that the divergence in
correlation lengths is symmetric between the superconducting and insulating
side of the transition due to the electric-magnetic
duality\cite{duality}. 
More precisely, the vortex proliferated insulator can be considered as
the bose-condensed state of vortices, and consequently superfluid
motion of vortices is possible.
Therefore, by choosing a superconducting junction sufficiently
close to SIT, it will be possible to detect 
the super-long-range proximity effect of the vortex-condensate
superfluid in the {\it dual} Josephson  (ISI) junction experiment\cite{AC}. 
\begin{figure}[t]
\centerline{\psfig{file=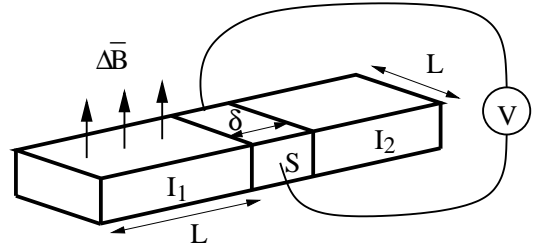,width=0.8\linewidth,angle=0}}
\caption{A dual Josephson junction experiment of
insulator-superconductor-insulator type.}
\label{fig:jcn}
\end{figure} 
In Fig. \ref{fig:jcn} we show an
ISI junction which
can be prepared by the same experimental technique used in Ref. \cite{Decca}.
We consider applying  an external magnetic field gradient across the
junction and subsequently letting it relax,
which will impose a  time-dependent phase difference ($\Delta \varphi$) in the
vortex-condensate order parameter ($\Phi$) across the junction.
The phase difference induces a dual Josephson current of magnetic vortices
($\vec{\cal J}$) across the superconducting junction. 
The vortex current leads to a transverse electric voltage difference
($\Delta V_{\rm tr}$) across the superconducting junction which can be subsequently measured.
As in the  superconducting proximity effect
experiment, we can imagine varying the length ($\delta$) of the junction to
probe the correlation length of the magnetic vortex field $\Phi_{\rm S}$ in the
superconducting junction. If the superconductor is in proximity to
the SIT, we expect a large vortex correlation length ($\xi_{\rm v}$),
analogous to the super-long-range proximity effect in the insulator
observed in Ref. \cite{Decca}. We show that a simple oscillatory solution to
the equation of dual Josephson currents can be obtained under certain 
conditions.

In this paper, we use the dual theory of vortices for a
Ginzburg-Landau description of vortex condensates in the insulating
state.
We first review the dual formulation\cite{SIT1,Kwon,dual,QED}. 
Throughout the paper we use the unit $\hbar =1$, and denote the
space-time three-vectors as $x= ({\bf x},t)$.
We begin with the effective phase-only action of a two-dimensional
(2D) zero-temperature 
superconductor near SIT coupled to the electromagnetic field $A_\mu$:
\begin{eqnarray}
&&S[\phi,A_\mu] ={1\over 2} \int d{\bf x}~dt \left\{
\rho_s\left[ (2A_0+\partial_t \phi)^2/c_s^2 -(2{\bf A}+\nabla\phi
)^2\right] \right. \nonumber \\
&&\quad +\partial_t \phi ~n_s/2{\}} +\sum_{k}\bigg{[} {|{\bf k}|\over 4\pi e^2}A_0^2 
-{({\bf k}^2c^2-\omega^2)d\over 8\pi e^2}{\bf A}^2\bigg{]}~,
\label{phiA}
\end{eqnarray}
where  $n_s$ is the superconducting
fermion density, $\rho_s$ is the 
superfluid phase stiffness defined as $\rho_s = n_s/4m_{\rm f}$,  
$m_{\rm f}$ is the effective fermion mass, 
$c_s $ is the  superfluid phonon velocity, and $d$ is
the thickness of the superconducting film. 
In a BCS superconductor, $c_s $ is related to the Fermi velocity $v_F$
by $c_s=v_F/\sqrt{2} $.
Here we chose the Coulomb gauge $\nabla\cdot {\bf A}=0$. 
 
Since we are interested in the effective theory of  magnetic
vortices,
we first separate $\partial_\mu \phi$ into $\partial_\mu \phi =\partial_\mu
\theta +{\cal A}_\mu$ where $\theta $ is the spin-wave type
Goldstone fluctuation and $ {\cal A}_\mu$ gives
the topologically
non-trivial phase gradients generated by  magnetic vortices, and 
 integrate
out both the $\theta$ and  $A_\mu$ fields to obtain the 
final effective theory of the ${\cal A}_\mu$.
The ${\cal A}_\mu$ fields  are related to the
magnetic vortex current density (${\cal J}^\mu $) by
 ${\cal J}^\mu =\epsilon^{\mu \nu \lambda}
\partial_\nu {\cal A}_\lambda$. 
Then we transform the dynamics of single-particle paths of magnetic vortices
into  the field dynamics by introducing a quantum vortex field $\Phi$ which is
coupled to the superfluid phonon field represented by a
 gauge field $G_\mu$. 
The resulting well-known dual form \cite{Kwon,dual} 
of the effective theory of vortices follows:
\begin{eqnarray}
&&S[\Phi,\Phi^*,G]=\int d^2x~dt~ [ |(\partial_t+2\pi iG_0)\Phi|^2 \label{Sd} \\
&&\quad 
 -v_{\rm v}^2|(\nabla + 2\pi i {\bf G})\Phi |^2 -m_{\rm v}^2v_{\rm
v}^4|\Phi|^2-u|\Phi|^4/2]+S[G_\mu] ~,
\nonumber \\
&&S[G_\mu]={1\over 2} \sum_{k}\left[{{\bf k}^2\over
{\cal K}_T(k)}|G_0|^2
 -{{\bf k}^2\over
{\cal K}_0(k)}|{\bf G}|^2\right] ~, \label{non-l}
\end{eqnarray}
where $S[G_\mu]$ is the action of the new U(1) gauge field $G_\mu$.
Here
${\cal K}_0 = \rho_s {\bf k}^2/({\bf k}^2c_s^2- \omega ^2+8\pi e^2\rho_s |{\bf
k}|)$
and 
${\cal K}_T\approx \rho_s {\bf
k}^2/(\lambda^{-2}+{\bf k}^2)$ where $\lambda $ is the penetration depth. 
We also included the repulsive potential ($u>0$) between vortices.
The magnetic vortex rest mass is obtained from the
energy of a pancake vortex  and given by
 $m_{\rm v}v_{\rm v}^2 \approx \rho_s\pi\ln \kappa $ 
where $\kappa = \lambda\Lambda_c $, where we assumed a vortex core size
of $\xi_0$ which provides the momentum (short-range) cutoff at
$\Lambda_c=\pi/\xi_0$  and the
frequency (short-time) cutoff at $v_F\Lambda_c$. Here we take $v_{\rm
v}\approx v_F$ on  physical grounds\cite{QED}.
From Eq. (\ref{Sd}), the vortex number density ($n_{\rm v}$) and the
vortex current  can be expressed by $\Phi$ as
\begin{eqnarray}
n_{\rm v}&=&{\cal J}_0/2\pi = {\rm Im}~2 \Phi^*\partial_0
\Phi +4\pi G_0|\Phi|^2  \label{J0} \\
\vec{\cal J}/2\pi &=& v_{\rm v}^2({\rm Im}~2 \Phi^*\nabla
\Phi +4\pi {\bf G}|\Phi|^2) ~. \label{Jx}
\end{eqnarray}
In addition to Eq. (\ref{Sd}), there is a term that couples the vortex
currents to the background superfluid density which acts as a dual
magnetic field for vortices, but we assume that this does not affect
the superfluidity of the quantum vortices in the insulating state
where the Cooper pairs are localized.

We now discuss the superconducting proximity effects in the insulating state.
 The action in Eq.\ (\ref{Sd}), if $t$ is analytically continued to the
imaginary time, can be viewed as the effective
3D Ginzburg-Landau functional of $\Phi$. 
Then  SIT  occurs upon Bose-condensation of the $\Phi$ fields,
 when $m_{\rm
v}^2<0$ so that $\langle \Phi \rangle \neq 0$ and 
the correlation function  in the mean-field approximation is
$\langle \Phi(x)\Phi^*(y) \rangle \approx |\Phi_0|^2 >0$,
 independent of $|x-y|$. 
The correlation function $\langle \Phi(x)\Phi^*(y) \rangle  $ can be
considered as the expectation value of the number of magnetic vortex
 paths that  connect $x$
and $y$, and therefore $\langle \Phi(x)\Phi^*(y) \rangle \approx
 |\Phi_0|^2 >0$ implies a finite probability
of arbitrarily long 2+1D vortex loops. Here we  demonstrate that the
 Bose-condensed $\Phi$ field introduces a new length scale related to
 the observed super-long correlation length\cite{Decca}.

We first derive an effective Ginzburg-Landau functional of
 the superconducting order parameter ($\Psi$) in the vortex-proliferated
 insulating state.
The derivation is straightforward once we identify the trajectories of
Cooper pairs with dual (electric) vortex lines in the dual
 theory\cite{dual}. 
The symmetry between the Cooper pairs and vortices
is clear from the form of Eqs. (\ref{phiA}) and (\ref{Sd}) which
have similar functional  forms in $\rho_s e^{i\phi}$ and $\Phi$
respectively, except for the topological term $\partial_t\phi n_s/2$
in Eq. (\ref{phiA})  which is absent in Eq. (\ref{Sd}) and action of
the gauge fields.
Besides, just as a Cooper pair picks up a phase factor of $2\pi $
upon encircling a vortex, a vortex picks up the same phase factor upon
encircling a Cooper pair. Therefore, in the superfluid state of vortices
(insulating state), a Cooper pair is the topological defect, playing
the role of a vortex. Thus, we can perform a dual transformation on
Eq. (\ref{Sd}) to obtain an effective theory of superconducting order 
parameter,  in the same way as the dual  theory of
vortices in Eqs.\ (\ref{Sd}) and (\ref{non-l}) is derived
from the original superconducting theory in Eq.\ (\ref{phiA}). The
result is following:
\begin{eqnarray}
&&S[\Psi, \Psi^*,A_\mu] = \int d^2 x ~dt~[|(\partial_t+
2iA_0)\Psi|^2/m_cv_c^2  \nonumber \\
&&\quad 
 -|(\nabla +  2i {\bf A})\Psi |^2/m_c -m_{\rm
c}v_c^2|\Psi|^2]+S[A_\mu]~,
\label{SPsi}\\
&&S[A_\mu]=\sum_k {1\over (2\pi)^2|\Phi_0|^2 v_{\rm v}^2}\left\{
\left[{\bf k}^2+2|\Phi_0|^2(2\pi)^2{\cal K}_0(k)\right]A_0^2
\right.\nonumber \\
&&\quad \left.
 -\left[v_{\rm v}^2{\bf k}^2-\omega^2+v_{\rm v}^2|\Phi_0|^2
(2\pi)^2{\cal K}_T(k)\right]{\bf A}^2 \right\}~.
\end{eqnarray}
Here we estimate that 
$m_cv_c^2=4\pi|\Phi_0|^2 v_{\rm v}^2 \ln(L/\xi_0)$ and 
take $v_c \approx v_{\rm v}$ on physical grounds, 
where $L$ is the linear dimension
of the sample.
Therefore, we may take the correlation length of the superconducting
order parameter from $\xi_c^{-1} \approx m_cv_c \approx
4\pi |\Phi_0|^2 v_{\rm v}\ln(L/\xi_0)$, and so
$\xi_c$ can be a very large number near  SIT where
$|\Phi_0|^2 $ is small.
This can explain the anomalous proximity effect observed in
Ref. \cite{Decca}, assuming that the insulating compound is in 
the  vortex-proliferated state.
The vortex-proliferated insulating state has a physically distinct property
of vortex superfluidity as shown below.

Next we consider the {\it dual} proximity effect
near SIT. We consider an ISI junction surrounded by a superconducting region so
that magnetic flux can be trapped inside. 
We assume that the vortex condensate density in I$_1$ and I$_2$ is 
$|\Phi_0|^2$. We imagine applying a magnetic
flux  density difference ($\Delta\bar{B}$) between I$_1$ and I$_2$
regions of the ISI junction and relaxing it 
at a certain time ($t=0$)
as shown in Fig. \ref{fig:jcn}, and consider the time evolution of the
magnetic field difference and the vortex supercurrent across the
S region. Here the difference in 
the total number of magnetic vortices is $\int d^2 x \langle \Delta{\cal J}_0
\rangle/2\pi = \Delta \bar{B}L^2/\phi_0$ where $\phi_0=hc/2e$ is a quantum of
magnetic flux in a superconductor and $L$ is the linear size of the
I$_1$ and I$_2$ regions. This condition can be implemented by adding a
Lagrange multiplier $\int dt~\mu_{\rm v} (t) 
\int d^2x [\Delta{\cal J}_0({\bf x},t) - 2\pi \Delta\bar{B}/\phi_0]$ to
Eq. (\ref{Sd}). 
The chemical potential difference $\mu_{\rm v}$ can be absorbed into the dual
electro-chemical potential difference 
$\bar{G}_0$ which is the sum of the chemical
potential and the {\it dual} scalar potential generated by the
presence of extra vortices introduced by the external magnetic field.
Ignoring the spatial fluctuations in $\bar{G}_0$,
we can absorb $\bar{G}_0$ into the phase difference of the field $\Phi$ by the
transformation $\Phi \rightarrow \Phi e^{i\Delta\varphi}$ where
\begin{equation}
\partial_t \Delta\varphi (t) =-2\pi\bar{G}_0(t)~ .
\label{phiG}
\end{equation}
The difference in the vortex number density  can be related to $\bar{G}_0$
from Eq.\ (\ref{J0}) as follows:
\begin{equation}
\Delta n_{\rm v}= \langle \Delta{\cal J}_0/2\pi
\rangle\approx 4\pi \bar{G}_0(t)|\Phi|^2~,
\label{vnumber}
\end{equation}
neglecting the time derivatives of the condensate $\Phi $.
The value of $\bar{G}_0$
is determined by the condition $ \Delta n_{\rm v}=
  \Delta\bar{B}/\phi_0 \approx 4\pi \bar{G}_0|\Phi|^2$, which gives
$\bar{G}_0 \approx \Delta \bar{B}/(4\pi|\Phi_0|^2 \phi_0)$. 
Inside the superconducting junction (S) of length $\delta$ and
width $L$, the $\Phi_{\rm S}$ field is phase-incoherent and does not 
have non-zero vacuum expectation value.
From Eq. (\ref{Sd}), ignoring non-linear effects, the semi-classical
$\Phi_S$ field minimizes the following action:
\begin{equation}
S= \int d^2x~dt[ |\partial_t\Phi|^2-v_{\rm v}^2|\nabla
\Phi|^2+v_{\rm v}^2 |\Phi/\xi_{\rm v}|^2]
~,
\label{diffE}
\end{equation}
with a correlation length of
$\xi_{\rm v}\approx 1/m_{\rm v}v_{\rm v}$.
If the field $\Phi_{\rm S} ({\bf x},t)$ 
is solved in the S region, we
can obtain the vortex number current 
[$I_{\rm v}=L{\cal J}_x({\bf x},t)/2\pi$] 
between I$_1$ and I$_2$ (in the $x$ direction) flowing through the S region. 
Then the rate of change of the vortex number density difference can be
related to the vortex current as follows: 
\begin{equation}
-\partial_t \Delta n_{\rm v}={\cal J}_x({\bf x},t)/2\pi L~.
\label{vcurrent}
\end{equation}

Below we show  that if we assume a nearly one-dimensional geometry of
the junction,
a simple solution to Eqs. (\ref{phiG}), (\ref{vnumber}),
(\ref{diffE}), and (\ref{vcurrent}) can be obtained under following conditions:
(a) The magnetic field difference ($\Delta\bar{B}$) is small so that 
$|\Phi|^2$ in the insulators can be approximately taken as a constant as a function
of $\Delta n_{\rm v}$. 
(b) Inside the superconducting junction region, 
the following conditions hold: 
$|\Phi_{\rm S}/\xi_{\rm v}|^2\ll |\partial_x \Phi_{\rm S}|^2$, which is
valid when $\delta \ll \xi_{\rm v}$, and $|\partial_t\Phi_{\rm
S}/v_F|^2\ll |\partial_x \Phi_{\rm S}|^2 $. In this case, from 
Eq.\ (\ref{diffE}),
the equation of motion for $\Phi_{\rm S} ({\bf x},t)$ is simply reduced
to a static Laplace equation: $\partial_x^2\Phi_{\rm S}= 0$.
(c) The phase difference remains small ($|\Delta \varphi | \ll 1$) due
to the small magnitude of the magnetic field difference ($\Delta\bar{B}$).
From condition (b), along with the boundary conditions
$\Phi_{\rm S}(0)=\Phi_{\rm 0} e^{i\Delta \varphi}$ and $\Phi_{\rm S}(\delta)=
\Phi_{\rm 0}$,
we can obtain the solution of the Laplace equation in the region 
S\cite{Likharev}:
\begin{equation}
\Phi_{\rm S} \approx \Phi _{\rm 0} [e^{i\Delta \varphi} (1-x/\delta)+
(x/\delta)]~.
\label{Phisol}
\end{equation}
Then one can show that there is  a magnetic vortex supercurrent 
through the region S as follows:
\begin{equation} {\cal J}_x \approx 2\pi v_{\rm v}^2
i(\Phi_{\rm S}\partial_x \Phi_{\rm S}^* -\Phi_{\rm S}^*\partial_x
\Phi_{\rm S} )\approx -\hat{x}J_{\rm
v} \sin  \Delta
\varphi ~,
\end{equation}
where $J_{\rm v} = 4\pi v_{\rm v}^2|\Phi_0|^2/\delta$. If $\delta $ exceeds the
correlation length
$ \xi_{\rm v} $, $J_{\rm v} $ will be exponentially suppressed in
$\delta/\xi_{\rm v}$\cite{Likharev}. 
From Eqs. (\ref{phiG}) and (\ref{vnumber}), and substituting the
vortex current obtained from Eq. (\ref{Phisol}) into
Eq. (\ref{vcurrent}),
we obtain the equation $\partial_t^2 \Delta\varphi
=-\omega_{\rm v}^2 \sin \Delta\varphi $, which reduces to
$\partial_t^2 \Delta\varphi \approx -\omega_{\rm v}^2\Delta\varphi $ from
condition (c) that $|\Delta \varphi |\ll 1$,
 where $\omega _{\rm v} = \sqrt{J_{\rm v}/4\pi L|\Phi_0|^2} =v_{\rm
v}/sqrt{ L\delta}$ is
the  natural frequency of the given ISI junction.
The solution in this case is 
\begin{equation}
\Delta\varphi (t) \approx \varphi_{\rm M} \sin
(\omega_{\rm v}t+\alpha)
\label{Jsol}
\end{equation} 
with $| \varphi_{\rm M} |\ll 1$. In the presence of such vortex
condensate phase oscillations,
we may detect the magnetic  vortex supercurrents by measuring the transverse
voltage difference across the region S (see Fig. \ref{fig:jcn}) 
which is induced by the vortex motion as follows:  
\begin{equation}
2e\Delta V_{\rm tr}= 
L{\cal J}_x\approx LJ_{\rm v} \varphi_{\rm M} \sin (\omega_{\rm
v}t+\alpha)~.
\label{Vtr}
\end{equation}
 
As the junction thickness $\delta $ is increased above $\xi _{\rm v}$,
the Laplace equation of $\varphi$ does not hold anymore in the region
S, and the relation $J_{\rm v}\propto 1/\delta$ breaks
down. Therefore, by varying the thickness $\delta $ and studying
$J_{\rm v}(\delta )$, the correlation length $\xi_{\rm v}$ can be
determined as the length scale of $\delta $ where $J_{\rm v}(\delta )$
deviates from  $J_{\rm v}\propto 1/\delta$. As we further underdope
the charge-carriers of the region S, we expect that $\xi_{\rm v}$ will
grow, resulting in a super-long-ranged dual proximity effect.

We are now able to make estimates of various quantities related to the
(dual) proximity effects in the  SIS (ISI) junctions, to see if we can
find a regime of physical parameters within experimental access. 
The information on one particular set of parameters is provided by
the experimental results of Decca {\it et
al.}\cite{Decca} on the SIS junction. The insulating junction (with
thickness $\delta_c$)
described by Eq. (\ref{SPsi}) leads to a Josephson current
$I_J = I_c \sin \Delta \varphi $ where $I_c =en_sL/2m_{\rm f}\delta_c$, if
$\delta_c \ll \xi_c$. In Ref. \cite{Decca}, $I_c \approx 2.6~ \mu$A with
$\delta_c \approx 45$ nm and $L\approx 200$ nm, which leads to $\rho_s =
n_s/4m_{\rm f} \approx 1.2$ meV. From this, using the 1D Josephson weak
link model\cite{Likharev}, one can estimate the magnetic  vortex
 rest energy $m_{\rm v}v_{\rm v}^2=\rho_s \pi \ln \kappa
 \approx 2.6\times 10^{-2}$ eV. 
Note that the above small magnitude of the superfluid phase stiffness
$\rho_s$ already indicates 
that the material is under strong vortex fluctuations and
the vortex mass $m_{\rm v}$ may have been significantly
renormalized since $\rho_s < 2$ meV\cite{Kwon},
 although we will  keep the bare mass for order of magnitude estimates.

Now we consider the ISI junction made of the same superconducting (S) and
 insulating (I$_1$ and I$_2$) regions as were used in Ref. \cite{Decca}.
Then the correlation length of the vortex fields in the region S
 is expected to be
$\xi_{\rm v}\approx (m_{\rm v}v_{\rm v})^{-1} \approx 5.1$ nm, which
 is a few times larger than the superconducting coherence length. 
We can enhance the correlation length $\xi_{\rm v}$ by further undoping
the charge-carrier density. Here we assume a junction of length $\delta
\approx 5.1$ nm and explore the possibility of the simple solution 
in Eq. (\ref{Jsol}). The magnitude of the fluctuations of $\Phi$ in
 time [condition (a)] cannot be checked within the
phenomenological theory presented here.
Condition (b) ($ \delta \ll \xi _{\rm v}$ and$ |\partial t \Phi_ {\rm
 S}/v_F| \ll |\partial_x  \Phi_ {\rm  S}|$ ) can be checked by
 comparing 
 $v_{\rm v}|\partial_x\Phi |/|\Phi |\sim v_{\rm v}/\delta
\approx $  10 THz
with the natural frequency 
$|\partial_t \Phi |/|\Phi |\sim \omega_{\rm v} 
= \sqrt{v_{\rm v}^2/L\delta}$ which is about  1 THz; we confirm that
condition (b) can be satisfied with the given parameters. 
Condition (c) ($|\Delta\varphi| \sim \Delta \bar B /(2\phi_0 \omega_{\rm
 v}|\Phi_0|^2) \ll 1$) gives the upper limit of the
magnetic field difference that
allows the solution in Eq. (\ref{Jsol}). From $\xi_c \approx 100$ nm
and $\xi_c^{-1} = 4\pi|\Phi_0|^2v_{\rm v}\ln (L/\xi_0)$, we find
that the upper limit of $ \Delta\bar{B}$ is $2.2\times 10^2$ Gauss. 
The upper limit of the transverse voltage in this case is
about $|\Delta V_{\rm tr}| \sim LJ_{\rm v}\varphi _{\rm M}/2e \sim 11$
mV, obtained from Eq. (\ref{Vtr}) and
the relation $(J_{\rm v}\xi_c)^{-1} = \delta \ln (L/\xi_0)/v_{\rm v}$.
Therefore, based on the information obtained by Decca {\it et
al.}\cite{Decca}, the proposed experiment of dual Josephson junction 
and the observation of the solution in Eq. (\ref{Jsol})
may be within the reach of experimental access.

We have shown that the anomalous proximity effect
in an underdoped insulating cuprate \cite{Decca} can be interpreted
in terms of vortex proliferation. 
We proposed that the dual counterpart of the proximity
effect can be  observed in an ISI junction experiment. We anticipate
that a similar super-long-range proximity effect will be observed in the dual
Josephson junction experiment as long as the junctions are prepared in
close proximity to the SIT. 

We thank H. D. Drew and V. M. Yakovenko for helpful comments.
This work was supported by the 
NSF DMR-9815094 and the Packard Foundation.


\begin{references}
\vspace{-5ex}

\bibitem{pseudogap} See, for example, 
F. Ronning {\it et al.}, Science {\bf 282}, 2067 (1998); A. Ino {\it
et al.}, \prb {\bf 62}, 4137 (2000).

\bibitem{SIT1} L. Balents, M. P. A. Fisher, and C. Nayak,
Int. J. Mod. Phys. B {\bf 12}, 1033 (1998); \prb {\bf 60},
1654 (1999). 

\bibitem{SIT2}S. Doniach and M. Inui, \prb {\bf 41}, 6668 (1991);
 S. Sachdev, {\it ibid.} {\bf 59}, 14054 (1999).

\bibitem{Kim} Y. B. Kim and Z. Wang, Europhys. Lett. {\bf 50}, 656 (2000).

\bibitem{Kwon}H.-J. Kwon,  \prb  {\bf  63}, 134511 (2001).

\bibitem{SO5}S. C. Zhang, Science {\bf 275}, 1089 (1997).

\bibitem{spinon}P. W. Anderson, Science {\bf 235}, 1196 (1987);
X.-G. Wen and P. A. Lee, \prl {\bf 76}, 503 (1996);
 M. Vojta, Y. Zhang, and S. Sachdev, 
Phys. Rev. B {\bf 62}, 6721 (2000); S. Chakravarty, R. B. Laughlin, D. K.
Morr, C. Nayak, {\it ibid.} {\bf 63}, 094503 (2001); C. Lannert,
M.P.A. Fisher, and T. Senthil, {\it ibid.} {\bf 63}, 134510 (2001).

\bibitem{stripes} J. Zaanen and O. Gunnarsson, {\it ibid.} {\bf 40}, 7391
(1989); O. Zachar, S. A. Kivelson, and V. J. Emery, {\it ibid.} {\bf
57}, 1422 (1995).

\bibitem{Decca}R. S. Decca {\it et al},  \prl {\bf 85}, 3708 (2000).

\bibitem{SO5Jo}E. Demler {\it et al.}, \prl {\bf 80}, 2917 (1998);
B. C. den Hertog, A. J. Berlinsky, and C. Kallin, \prb {\bf 59},
R11645 (1999).

\bibitem{Corson} J. Corson {\it et al.}, Nature (London) {\bf 398},
221 (1999).

\bibitem{Xu} Z. A. Xu {\it et al.}, {\it ibid.} {\bf 406}, 486  (2000).

\bibitem{Emery} V.J. Emery and S.A. Kivelson, {\it ibid.} {\bf 374},
434 (1995).

\bibitem{Chiaverini} J. Chiaverini {\it et al.}, cond-mat/0007479.

\bibitem{duality}D. Shahar {\it et al.}, Science {\bf 274}, 589
(1996); Y. Aharonov and A. Casher, \prl {\bf 53}, 319 (1984);
B. J. van Wees, {\it ibid.} {\bf 65}, 255 (1990);
W. J. Elion, J. J. Wachters, L. L. Sohn, and J. E. Mooij,
{\it ibid.} {\bf 71}, 2311 (1993).

\bibitem{AC} Dual Josephson junctions of a Corbino disk geometry
 were studied in A. Vourdas, Europhys. Lett. {\bf 48}, 201 (1999).

\bibitem{dual}See, for example, M.P.A. Fisher and D.H. Lee, \prb
{\bf 39}, 2756 (1989); H. Kleinert, {\it Gauge Fields in Condensed
Matter} (World Scientific, Singapore, 1989) Vol.1.

\bibitem{QED}
V. N. Popov,Zh. Eksp. Toer. Fiz. {\bf 64}, 672 (1973) 
[Sov. Phys. JETP {\bf 37}, 341 (1973)]; 
D. P. Arovas and J. A. Freire, \prb {\bf 55}, 1068 (1997).

\bibitem{Likharev} L. G. Aslamazov and
A. I. Larkin, Zh. Eksp. Teor. Fiz. Pis'ma Red. {\bf 9}, 150 (1969)
 [JETP Lett. {\bf 9}, 87 (1969)]; K. K. Likharev, \rmp {\bf 51}, 101 (1979).


\end{references}
\end{document}